# Spin-layer coupling in altermagnets multilayer: a design principle for spintronics

Jianke Tian,[1] Jia Li,[2,*], Hengbo Liu,[1] Yan Li,[1] Ze Liu,[1] Linyang Li,[1] Jun Li,[1] Guodong Liu,[1] Junjie Shi[3]

[1]*School of Science, Hebei University of Technology, Tianjin 300401, People's Republic of China*

[2]*College of Science, Civil Aviation University of China, Tianjin 300300, People's Republic of China*

[3]*State Key Laboratory for Artificial Microstructures and Mesoscopic Physics, School of Physics, Peking University Yangtze Delta Institute of Optoelectronics, Peking University, 5 Yiheyuan Street, Beijing, 100871, People's Republic of China.*

The discovery of collinear symmetric-compensated altermagnets (AM) with intrinsic spin splitting provides a route towards energy-efficient and ultrafast device applications. Here, using first-principles calculations and symmetry analysis, we propose a series of AM $Cr_2SX$ (X=O, S, Se) monolayer and explore the spin splitting in $Cr_2SX$ multilayer. A general design principle for realizing the spin-layer coupling in odd/even-layer is mapped out based on the comprehensive analysis of spin group symmetry. The spin splitting behavior related with the $\mathcal{M}_z\mathcal{U}t$, $\mathcal{M}_z$ and $\mathcal{M}_L$ symmetries in AM multilayer can be significantly modulated by magnetic orders, crystal symmetry and external perpendicular gate field ($E_z$). Due to the spin-compensated bands of sublayers linked by overall $\mathcal{M}_z$ and interlayers $\mathcal{M}_L$ symmetries, the $Cr_2S_2$ odd-layer exhibits the unique coexistence of spin splitting and spin degeneracy at high symmetric paths and X/Y valley, respectively. Furthermore,

---

*Corresponding author. *E-mail address*: jiali@hebut.edu.cn (Jia Li).

owing to the higher priority of overall $\mathcal{M}_L$ symmetry compared to interlayers $\mathcal{M}_L$ symmetry in AM even-layer, the spin-layer coupling of AM multilayer shows strong odd/even-layer dependence. Our work not only offer a new direction for manipulating spin splitting, but also greatly enrich the research on AM monolayer and multilayer.

## I. INTRODUCTION

Exploring new degrees of freedom (DOF) and finding an efficient and precise control of spin are crucial to the fabrication of advanced spintronic devices [1,2]. Collinear antiferromagnetism (AFM) systems are robust to external magnetic perturbation due to compensated magnetic order [3-5]. However, the missing spin splitting in band structures of AFM systems makes people mainly focus on materials with ferromagnetism (FM) which are susceptible to thermal/magnetic perturbation [6-10]. Recently, the spin splitting has been found in the third elementary type magnetic phases without spin-orbit coupling (SOC), i.e. AM, in addition to the conventional FM and AFM [11-18]. The spin splitting originates from the exclusively distinct spin-symmetry, that is, the opposite-spin sublattices are connected by a real-space rotation rather than translation or inversion transformation, which protects an antiferromagnetic-like vanishing net magnetization [11,13]. The achievement of spin splitting in AM without SOC and the valley protected by lattice symmetry facilitates their practical applications for spintronic, valleytronic and thermal transport devices [16-19]. Several bulk AM systems, such as $RuO_2$ [17,18], $FeF_2$ [20], MnTe [21], AM topological systems $Fe_2WTe_2$, $Fe_2MoZ_4$ (Z=S, Se, Te) [22], monolayer AM CrO [23], $V_2Se_2O$ [24] and some $GdFeO_3$-type perovskites [25] have been predicted.

Two-dimensional (2D) van der Waals (vdW) systems bring novel layer DOF due to the weak interlayer vdW interaction, and flexible vdW stacking releases highly tunable spin and valley properties [26-29]. For instance, the layer and valley DOFs can be coupled to form the layer-polarized anomalous Hall effect (LP-AHE) in $MnBi_2Te_4$ [30] and $VSi_2P_4$ [31] bilayers, and the layer spin Hall effect (LSHE) can be operated by sliding ferroelectricity in $MoTe_2$ and $MoSi_2P_4$ bilayers [32]. However, the manipulation of spin/valley splitting in AM multilayer is still lacking, although various conventional FM and AFM bilayers have been investigated [33,34]. In the view of symmetry, the spin/valley properties in AM multilayer may be regulated via interlayer coupling [35]. Moreover, unlike a monolayered nanostructure with only in-plane wave vector $k$, for the AM multilayer, the out-of-plane component of layer-dependent pseudospin corresponds to the electric dipole moment, which indicates that the perpendicular gate field ($E_z$) could directly couple to the layer DOF and then may lead to indirect control of spin-layer coupling via certain mechanisms [36,37]. Therefore, it is crucial for spintronics to explore the spin-layer coupling mechanism and manipulative methods in AM multilayer [38].

In this work, based on the first-principles calculations, we predict a series of 2D AM $Cr_2SX$ (X=O, S, Se) monolayer, and give the design principle of spin-layer coupling in $Cr_2SX$ multilayer. First, we theoretically proposed that $Cr_2SX$ monolayer shows a pair of spin-polarized valleys, and the states of conduction and valence bands at X/Y valley are occupied by opposite spin-channels. Then, we demonstrated that the spin-layer coupling is significantly affected by spin group symmetry, and the spin splitting can be highly tuned through a perpendicular gate field ($E_z$). Finally, based on the view of symmetry, we reveal the coexistence of spin splitting and spin degeneracy in $Cr_2S_2$ odd-layer, emphasizing the significant role of overall/interlayer $\mathcal{M}_L$ symmetry in spin-layer dependent coupling. Our theoretical investigation provide clear physical insights and a series of practical material platforms for the realization of intriguing coupling between spin/valley and layer physics.

## II. COMPUTATIONAL DETAILS

All structural optimization and electronic structure calculations are performed on the basis of density functional theory (DFT) by employing the Vienna *ab initio* simulation package (VASP), and the projected augmented wave (PAW) method [39,40]. The kinetic energy cutoff and total energy convergence criterion is set to 500 eV and $10^{-6}$ eV, respectively. Atomic positions are fully relaxed until the maximum force on each atom is less than 0.01 eV/Å. A 20 Å vacuum layer in the z direction is used to eliminate interactions between adjacent layers. The Monkhorst−Pack k-points were sampled in the 2D Brillouin zone using a Γ-centered 11×11×1 k-grid. To better describe the strong correlation effect for Cr atoms, a Hubbard correction $U_{eff}$ is employed within the rotationally invariant approach [33,41]. The SOC effect is included in the calculations [42]. The phonon dispersion is based on a 4×4×1 supercell by using the PHONOPY code [43]. The *ab initio* molecular dynamic (AIMD) simulations adopt the *NVT* ensemble based on the Nosé-Hothermostat [44]. To describe the interlayer interaction of bilayers, the DFT-D3 method is employed to deal with van der Waals interaction [45].

### III. RESULTS AND DISCUSSIONS

#### A. Design principle of spin-layer coupling for AM multilayer

We firstly consider a 2D collinear AFM-Neel system, in which the absence of long-range FM ordering leads to the emergence of fully spin-compensated bands. In fact, the energy eigenvalue of $E_↑(k) = E_↓(k)$ for AFM-0 in Fig. 1(a) originates from the protection of $\mathcal{PT}$ symmetry contributed by antiparallel magnetic moments [46]. The $\mathcal{P}$ operation reverses the vector *k* to produce $\mathcal{P}E_↑(k) = E_↑(-k)$, and $\mathcal{T}$ operation reverses both *k* and spin σ, i.e. $\mathcal{T}E_↑(k) = E_↓(-k)$. Thus, the energy eigenvalue of AFM system protected by $\mathcal{PT}t$ symmetry is $\mathcal{PT}t E_↑(k) = E_↓(k)$, where *t* is translation operation with $tE_↑(k) = E_↑(k)$. For FM systems, the breaking of $\mathcal{T}$ symmetry causes the spin-splitting $E_↑(k) ≠ E_↓(-k)$, and the corresponding electronic band structures for K/K' valley of valleytronic materials can be classified as bipolar-semiconductor band (FM-0) and half-semiconductor band (FM-1) in Fig. 1(a). When the SOC effect is negligible, the spin space is completely independent of the real space, and the spin-reversal operation $\mathcal{U}$ ($\mathcal{U}E_↑(k) = E_↓(k)$) that reverses the signs of the spin and magnetic moments can be introduced.

Based on the above analysis, the occurrence of SOC-unrelated spin splitting requires the violations of $\mathcal{U}t$ and $\mathcal{PT}t$ symmetries. However, for a AM monolayer, the lacking of out-plane wave vector *k* remains the energy eigenvalues constant under planar mirror reflection $\mathcal{M}_z$ operation $\mathcal{M}_z E_\uparrow(k) = E_\uparrow(k)$. Therefore, the spin splitting is not allowed to exist in AM monolayer when opposite-spin sublattices can be transposed onto each other by four joint symmetries $\mathcal{U}t$, $\mathcal{PT}t$, $\mathcal{M}_z\mathcal{U}t$ or $\mathcal{M}_z\mathcal{PT}t$ [13,15,35,47]. Interestingly, for opposite-spin sublattices of adjacent sublayer in AM multilayer, the symmetric operations results of $\mathcal{PT}t$ and $\mathcal{M}_z\mathcal{U}t$ symmetries are equivalent, and $\mathcal{U}t$ and $\mathcal{M}_z\mathcal{PT}t$ symmetries are naturally broken, so we only need to discuss $\mathcal{M}_z\mathcal{U}t$ symmetry in adjacent sublayer. In addition, we further define the mirror reflection symmetry $\mathcal{M}_L$ considering magnetic configuration of Cr atoms in AM multilayer, and demonstrate the significant influence of $\mathcal{M}_L$ symmetry in spin degeneracy.

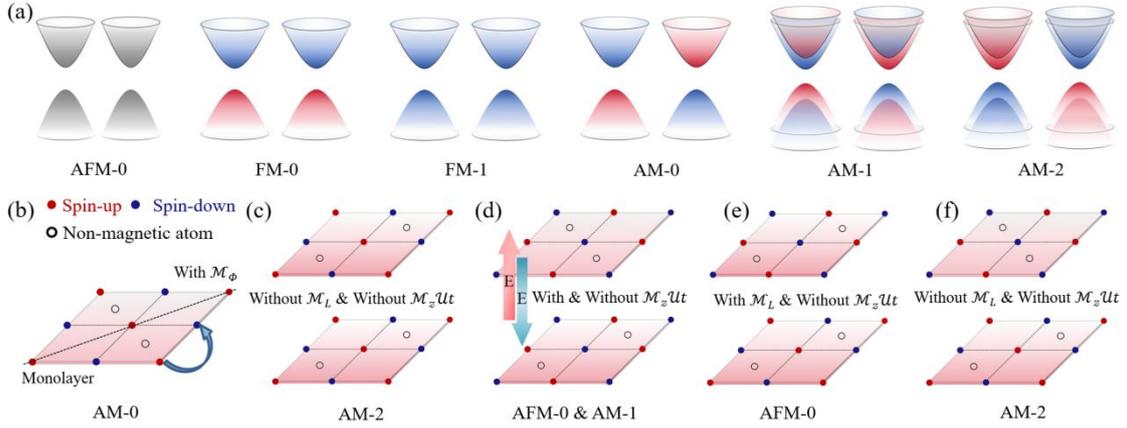

FIG. 1. (a) The schematic diagrams of different electronic band structures for 2D AFM, FM and AM semiconductors. The colors gray, red and blue represent the non spin-splitting, spin-up and spin-down bands, respectively. (b) The AM monolayer crystal structure with $\mathcal{M}_\Phi$ symmetry. (c) The schematic diagrams of AM bilayer without $\mathcal{M}_L$ and $\mathcal{M}_z\mathcal{U}t$ symmetries. (d) The AM bilayer with (without) $\mathcal{M}_z\mathcal{U}t$ symmetry in non-applying (applying) the external electric field. The vertical red/blue arrow represents the direction of external electric field. (e) Based on the distribution of magnetic atoms, the bilayer system with (without) $\mathcal{M}_L$ ($\mathcal{M}_z\mathcal{U}t$) symmetry. (f) In contrast to (e), the bilayer system without $\mathcal{M}_L$ and $\mathcal{M}_z\mathcal{U}t$ symmetries.

We first theoretically analyze a 2D tetragonal AM nanostructure in Fig. 1(b). The special spin space group that break four joint symmetries of AM monolayer causes the spin splitting, and the corresponding band structure AM-0 is shown in Fig.1(a). Notably, the in-plane diagonal mirror symmetry $\mathcal{M}_\Phi$ associated with the energy valleys (X/Y valley) near the Fermi level is preserved, suggesting that uniaxial strain can induce spin-valley polarization. Subsequently, we build a AM bilayer without

$\mathcal{M}_z\mathcal{U}t$/ $\mathcal{M}_L$ symmetries due to the distribution of non-magnetic/magnetic atoms, in which the magnetic atoms in the top and bottom sublayers have the same spin (in Fig. 1(c)). The electronic band structure (AM-2) of this system at X/Y valley can be regarded as the combination of the electronic band structure of two AM monolayer. In contrast, we consider another AM bilayer with interlayer non-magnetic atoms in a staggered arrangement (in Fig. 1(d)), where the magnetic configurations of sublayers are also same, reflecting that the $\mathcal{M}_L$ symmetry is broken. However, the spin sublattices of top and bottom layers can be correlated by $\mathcal{M}_z\mathcal{U}t$ symmetry, indicating that two sets of electronic band structures recombine and lead to complete spin degeneracy (like AFM-0 in Fig. 1(a)). By applying an external electric field in the $z$ direction to break $\mathcal{M}_z\mathcal{U}t$ symmetry, the AM-1 spin splitting can be found in this system. Notably, the X and Y valleys are still protected by crystal symmetry $\mathcal{M}_\Phi$ with no valley splitting, i.e. $E_{Xc(v)} = E_{Yc(v)}$. Furthermore, two other AM bilayers are considered and one crystal structure is shown in Fig. 1(e), which is similar to that in Fig. 1(c), except that the magnetic configuration of top and bottom sublayers are opposite. Although the $\mathcal{M}_z\mathcal{U}t$ symmetry of the system is broken, the existence of $\mathcal{M}_L$ symmetry makes the spin degenerate (AFM-0 in Fig. 1(a)). On the basis of AM bilayer in Fig. 1(d), we construct a system without $\mathcal{M}_L$ and $\mathcal{M}_z\mathcal{U}t$ symmetries using the opposite magnetic configuration of sublayers as shown in Fig. 1(f), and this system is expected to exhibit spin splitting (AM-2) independent of magnetic order.

Specifically, the $\mathcal{M}_z\mathcal{U}t$ and $\mathcal{M}_L$ symmetries of sublattices connecting the top and bottom sublayers in AM bilayer dominate the spin splitting in momentum space. A natural thought is whether the design principles of spin-layer coupling based on AM bilayer can be generalized to AM multilayer. Although the crystal structure of AM multilayer are complex, it is expected that the interlayer joint symmetries and $\mathcal{M}_\Phi$ symmetry are still the key to induce spin/valley splitting. In AM multilayer, the spin splitting is predicted to be dominated by $\mathcal{M}_z\mathcal{U}\,t$ symmetry of opposite-spin sublattices and $\mathcal{M}_z$ ($\mathcal{M}_L$) symmetries of crystal structure and magnetic configuration in adjacent sublayers. Therefore, the layer DOF that affect the joint symmetries is coupled with spin splitting to form the spin-layer locking. To verify the design principle of spin-layer coupling in AM multilayer, we predict the AM $Cr_2SX$ monolayer and multilayer with different crystal structures and magnetic configurations.

## B. The tunable spin splitting in $Cr_2SX$ monolayer and bilayer

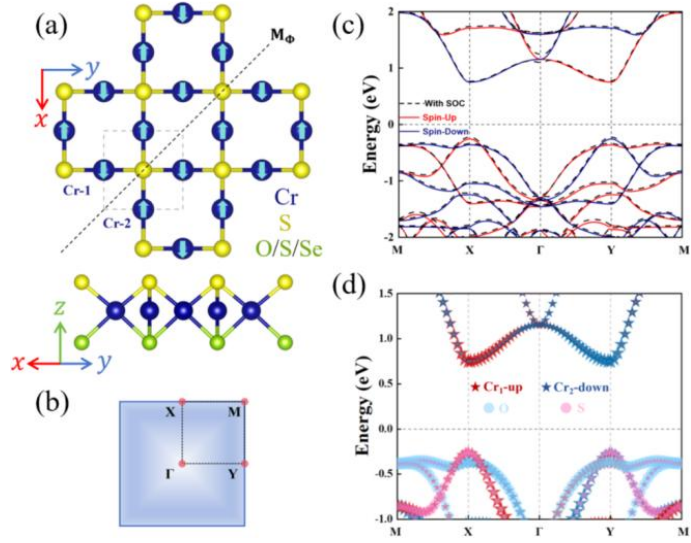

FIG. 2. (a) The top and side views of $Cr_2SX$ monolayer. (b) The first Brillouin zone with high symmetry points. (c) The electronic band structure of $Cr_2SO$ monolayer. (d) Spin-projected band structure of different atoms in $Cr_2SO$ monolayer.

In addition to elucidating the design principle of spin-layer coupling, it is equivalently important to identify candidate AM materials. Here, we propose a series of 2D AM $Cr_2SX$ (X = O, S, Se) materials as illustrated in Fig. 1(b). The monolayer $Cr_2SX$ has a square lattice structure, which consists of three atomic layers in the sequence of S-Cr-X (see Fig. 2(a)). $Cr_2S_2$ monolayer crystallizes in the *P*4/*mmm* space group (No. 123), and $Cr_2SY$ (Y=O, Se) monolayer possesses *P*4*mm* space group (No. 99). For $Cr_2SY$ monolayer, the symmetry operations for *P*4/*mmm* space group contain $\mathcal{P}$, $C_4$ rotation, $\mathcal{M}_\Phi$, and $\mathcal{M}_{x/y}$ mirror, while $Cr_2S_2$ monolayer has additional $\mathcal{M}_z$ symmetry. Four magnetic configurations of a 2×2×1 $Cr_2SX$ supercell (see Fig. S1 in the Supplemental Material) are constructed and the magnetic ground state is determined by calculation. Table S1 shows the relative energies between FM and AFM configurations when AFM-Neel configuration energy is set to zero, indicating that the AFM-Neel state is the ground state of $Cr_2SX$. The AFM-Neel state is associated with AM, enabling spin splitting in conventional collinear AFM.

Fig. S2 presents the phonon spectrum of $Cr_2SX$ for AFM-Neel case by using GGA+$U$ calculation, where there is no imaginary frequency, which confirms the structural stability. The thermal stability of $Cr_2SX$ was also verified by AIMD with small energy fluctuations and the preservation of structural integrity. The elastic constants (Table S1) satisfy the Born-Huang criteria [48], i.e. $C_{11} > 0$ and $C_{11}>|C_{12}|$, indicating the mechanical stability of $Cr_2SX$. In summary, we predict a series of highly stable $Cr_2SX$ in the AFM-Neel ground state, and believe that $Cr_2SX$ can be realized experimentally in the future.

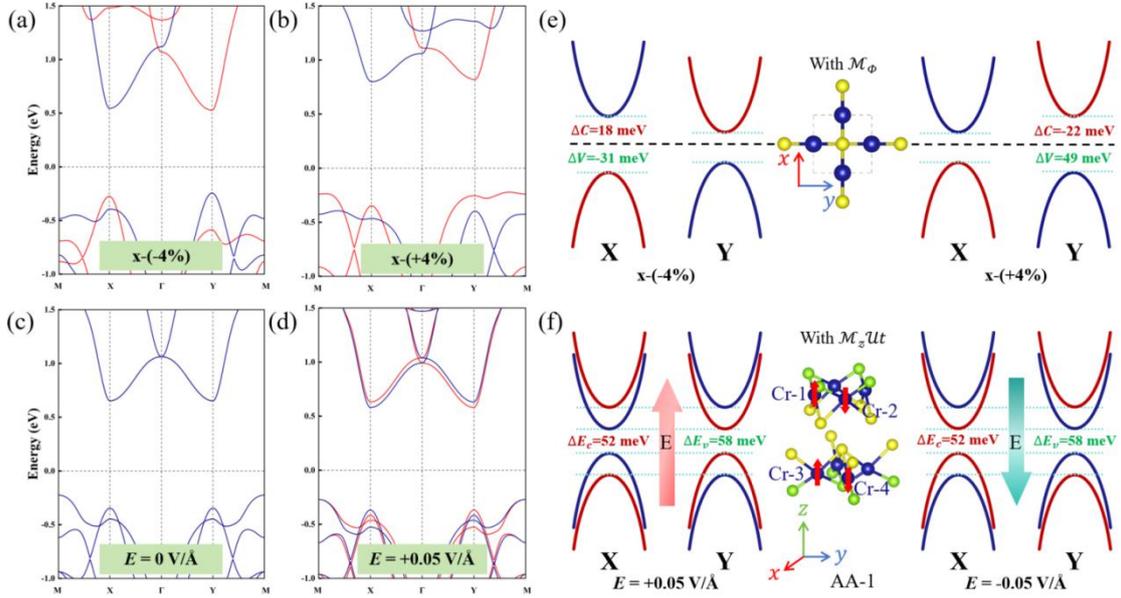

FIG. 3. The electronic band structures of $Cr_2SO$ under (a) -4% and (b) 4% uniaxial strain along x direction. The electronic band structures of AA-1 stacking under the external electric field of (c) 0 V/Å and (d) +0.05 V/Å. (e) Schematic of valley splitting induced by uniaxial strain. (f) The atomic structure of AA-1 stacking, and the schematic of spin splitting induced by an electric field. The red and blue dashed lines represent the spin-up and spin-down bands, respectively.

We focus on the electronic band structures of $Cr_2SO$ monolayer in the main text as the most representative one, and the electronic band structures of $Cr_2S_2$ and $Cr_2SSe$ monolayer are shown in Figs. S3 and S4. It can be found that $Cr_2SO$ monolayer shows semiconducting character with a 1.01 eV direct band gap (see Fig. 2(c)). According to the AM monolayer in Fig. 1(b), the opposite-spin sublattices in monolayer are connected by $\mathcal{PTC}_{4z}$ symmetry, which leads to the existence of AM that gives the

X/Y valley the opposite spins (like AM-0 in Fig. 1(a)). Near the Fermi level, the valence band maximum (VBM) and conduction band minimum (CBM) of $Cr_2SX$ monolayer around X or Y valleys are occupied by different Cr atoms with opposite spins as shown in Figs. 2(d) and S3. Since $Cr_2SO$ has negligible SOC effect, it can be considered as an AM without SOC effect (in Fig. 2(c)).

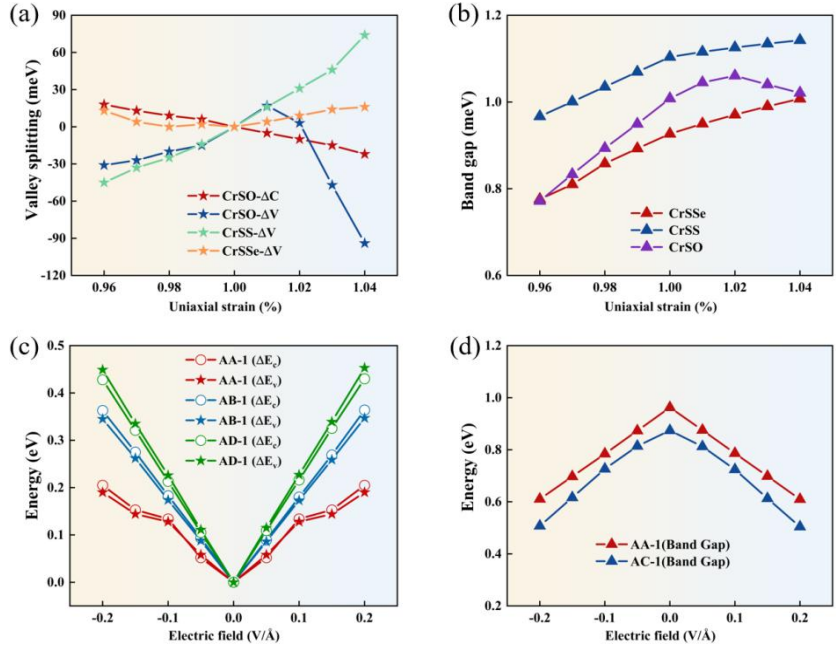

FIG. 4. Strain-controlled (a) valley splitting generated at the valence band ($\Delta V$) and conduction band ($\Delta C$), and (b) band gaps in $Cr_2SX$ monolayer. (c) The spin-splitting at the valence band ($\Delta E_v$) and conduction band ($\Delta E_c$), and (d) band gaps of AA-1 and AC-1 stackings modulated by external electric fields.

Generally, valleys splitting are formed in transition-metal dichalcogenides (TMDs) with strong SOC and the breaking of $\mathcal{PT}$ symmetry [49,50], however, the presence of valleys splitting in $Cr_2SX$ monolayer requires uniaxial strain to break the $\mathcal{M}_\Phi$ symmetry of crystal structure. In fact, the two degenerate X/Y valley which can be broken and show valley splitting are connected by crystal $\mathcal{M}_\Phi$ symmetry (C-paired valley) or by $\mathcal{T}$ symmetry (T-paired valley) [24,51]. To induce spin-valley polarization of C-paired valley in AM, an experimentally feasible approach is applying uniaxial strain field so as to breaks the $\mathcal{M}_\Phi$ symmetry. As shown in Figs. 3(e) and S3, we use $a/a_0$ (0.96-1.04) along the x direction to simulate the uniaxial strain, where a and $a_0$ are strained and unstrained lattice constants, respectively. The valley splitting of $Cr_2SX$ monolayer is defined as the energy difference $\Delta C$ and $\Delta V$ between X and Y valleys at CBM and VBM, respectively, where $\Delta C(V) = E_{Xc(v)} - E_{Yc(v)}$. The evolution of related valley splitting and band gap for $Cr_2SX$ monolayer as a function of uniaxial strain are plotted in Figs. 4(a) and 4(b). It is clearly seen that both compressive and tensile strains can induce valley splitting between X and Y valleys, and the valley splitting up to 74 meV in $Cr_2S_2$ induced by 4% tensile strains exceeds most ferrovalley materials, such as $LaBr_2$ (33 meV) [52], $YCl_2$ (22 meV) [53] and CrOBr (44 meV) [54]. In addition, the electronic band structures of $Cr_2SX$ monolayer regulated by strong correlation effect of Cr atoms $U_{eff}$ are shown in Fig. S3.

For $Cr_2SO$ bilayer, we consider four stackings (AA, AB, AC and AD) with different magnetic orders, and the relevant crystal structures and electronic band structures are shown in Figs. 3(f), S5 and S6. To facilitate the discussion for the symmetry, the magnetic/non-magnetic atoms in top layer are directly set above the magnetic/non-magnetic atoms of the bottom layer in all stackings except the AA stacking. The AA-1 and AA-2 stackings correspond to the AM bilayer models shown in Fig. 1(d) and Fig. 1(f), respectively, with the staggered arrangement of non-magnetic atoms. According to different magnetic configurations of sublayers, the AB, AC and AD stackings are described by the models in Fig. 1(c) and 1(e). Notably, AB and AD stackings possess $\mathcal{M}_z$ symmetry, while AC stacking has no $\mathcal{M}_z$ symmetry. Considering the total energy of four stackings with different magnetic configurations, it can be found that AA-2, AB-1, AC-1 and AD-1 stackings (i.e. the corresponding magnetic atoms of adjacent sublayers have opposite-spin) have the lowest total energy. Obviously, the AA-1 (AB-1, AD-1) stacking are coupled by $\mathcal{M}_z\mathcal{U}t$ ($\mathcal{M}_L$) symmetry, and the corresponding AM bilayer model is shown in Fig. 1(d) (Fig. 1(e)). As shown in Fig. 3 and S5, the AFM-0 spin degeneracy can be found in AA-1, AB-1, AD-1 stackings based on the spin-layer coupling principle. It is evident that in AM bilayer, both $\mathcal{M}_L$ and $\mathcal{M}_z\mathcal{U}t$ symmetries must be broken simultaneously for spin splitting to occur.

Analogous to the valley splitting, we define the spin splitting of X/Y valley at

VBM (CBM) as $\Delta E_{v(c)}=|E_{v(c)}^{\uparrow}-E_{v(c)}^{\downarrow}|$. Taking the AA-1 stacking corresponding to the model shown in Fig. 1(d) as an example, the external electric field ($\pm 0.05$ eV/Å) breaks the $\mathcal{M}_z\mathcal{U}t$ symmetry that connects sublayers, and the AM-1 spin splitting $\Delta E_v$ and $\Delta E_c$ of 58 and 52 meV occurs in X and Y valleys (in Fig. 3(f)), respectively. According to Fig. S7, the spin band contribution of Cr atoms in sublayers is similar to that of monolayer $Cr_2SO$. In addition, the spin splitting of AA-1, AB-1, AD-1 stackings increases linearly with the increase of external electric field (in Fig. 4(c)), which indicates the regulation of spin-layer coupling by external electric fields. As shown in Fig. S5, we next focus on the AA-2, AB-2, AC-1/2, and AD-2 stackings. It can be found that the spin splitting occurs in these stackings due to the absence of $\mathcal{M}_z\mathcal{U}t$ and $\mathcal{M}_L$ symmetries, and the AA-2, AB-2, AC-2 and AD-2 stackings have the corresponding AM-2-type spin splitting. It is noteworthy that the AC-1 stacking crystal structure lacks $\mathcal{M}_z$ symmetry, which leads to the breaking of $\mathcal{M}_L$ symmetry and the AM-1 spin splitting. In contrast with the AFM-0 spin degeneracy of AB-1 and AD-1 stackings with $\mathcal{M}_z$ and $\mathcal{M}_L$ symmetries, the breaking of $\mathcal{M}_z$ symmetry in AC-1 stacking plays a critical role in spin-layer coupling.

### C. The spin-layer dependent coupling in $Cr_2SX$ odd/even-layer

In this section, we demonstrate the anticipated spin-layer dependent coupling properties in $Cr_2S_2$ trilayer/pentalayer and $Cr_2SSe$ quadlayer, including spin-layer locking, layer-dependent coupling and valley-contrasting properties. As shown in Fig. 5(a)-(c), we built the different magnetic configurations (type-1 and type-2) in $Cr_2S_2$ trilayer satcking according to the design principles. Under the type-1 (type-2) configurations, the spin magnetic moments of Cr atoms are opposite (same) at different sublayers in $z$ direction, and the type-1 configuration is the most energetically favored for $Cr_2S_2$ trilayer satcking ($\Delta E = 11.0$ meV). In contrast to Janus $Cr_2SO$ and $Cr_2SSe$ materials, the crystal structure of $Cr_2S_2$ monolayer has $\mathcal{M}_z$ symmetry. The type-1 $Cr_2S_2$ trilayer has $\mathcal{M}_z$ symmetry, and two adjacent sublayers are linked by interlayer $\mathcal{M}_L$ symmetry, which should results in the spin degeneracy at the X/Y valley. However, the $\mathcal{M}_L$ symmetry of entire type-1 $Cr_2S_2$ trilayer is naturally broken, suggesting that the complete AFM-0 spin degeneracy does not exist in this trilayer system. As shown in Fig. 5(b), the resulting spin splitting is located near the high symmetric point Γ and high symmetric paths Γ-X/Y-M. The coexistence of spin splitting and spin degeneracy also occurs in type-1 $Cr_2S_2$ pentalayer (in Fig. S8). Subsequently, in type-2 $Cr_2S_2$ trilayer, we construct the opposite magnetic configuration between Cr-1, Cr-2, Cr-3 and Cr-4, Cr-5, Cr-6 atoms to further break the adjacent sublayer $\mathcal{M}_L$ symmetry in type-1 configuration. In Fig. 5(c), the electronic band structure of type-2 $Cr_2S_2$ trilayer, which lacks overall/interlayer $\mathcal{M}_L$ symmetry, exhibits the complete AM-2 spin splitting. The $Cr_2S_2$ pentalayer follows the same spin-layer coupling design principle as $Cr_2S_2$ trilayer, and its electronic band structure is presented in Fig. S8(g)-(i).

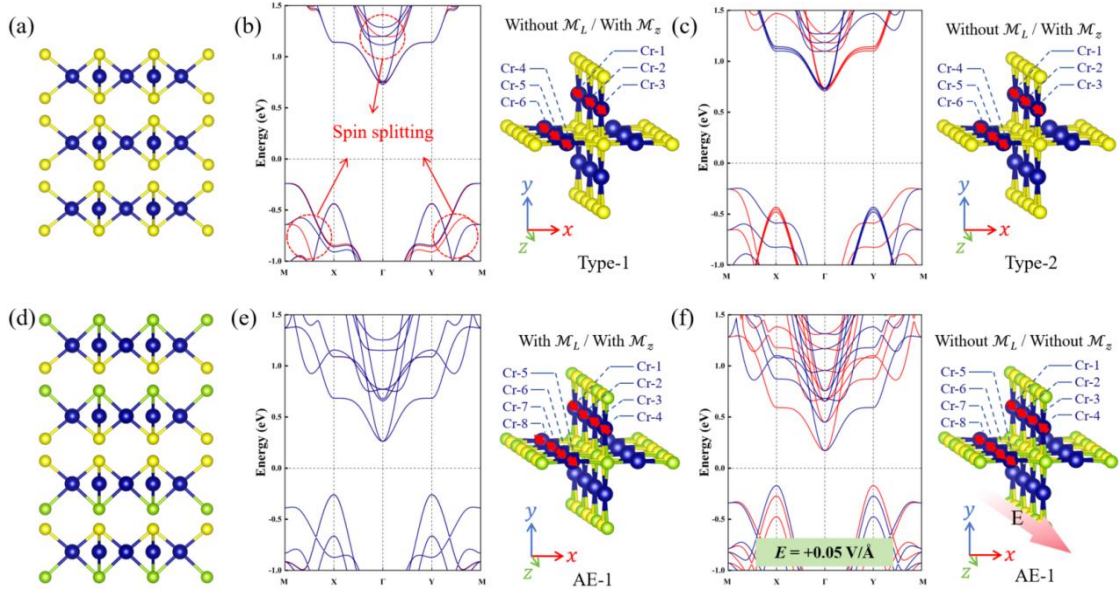

FIG. 5. (a) The side view of $Cr_2S_2$ trilayer. (b), (c) The different magnetic configurations and electronic band structures of $Cr_2S_2$ trilayer. (d) The side view of $Cr_2SSe$ quadlayer. The magnetic configuration and electronic band structure of $Cr_2SSe$ quadlayer (e) with and (f) without external electric field.

For AM odd-layer system, it also should be emphasized that the overall $\mathcal{M}_z$ symmetry, in addition to overall/interlayer $\mathcal{M}_L$ symmetry, is important for the coexistence of spin splitting and spin degeneracy, which can be demonstrated by the electronic band structures of $Cr_2SSe$ trilayer (in Figs. S8(d)-(f)), where the breaking of overall $\mathcal{M}_z$ symmetry leads to the complete spin splitting. In addition, similar to the

AM bilayer, the application of external electric field (+0.05 V/Å) can break the $\mathcal{M}_z$ and $\mathcal{M}_L$ symmetries of AM odd-layer and induce the complete spin splitting as shown in Fig. S8(b). Overall, in the above discussion, we have verified the symmetry conditions (i.e. the presence of $\mathcal{M}_z$ symmetry for entire system and interlayer $\mathcal{M}_L$ symmetry, and the absence of $\mathcal{M}_L$ symmetry for entire system) required for the coexistence of spin splitting and spin degeneracy.

In analogy to $Cr_2S_2$ odd-layer, we next discuss the property of $Cr_2SSe$ quadlayer crystal structure with overall $\mathcal{M}_z$ symmetry, and three magnetic configurations (AE-1, AE-2 and AE-3) are considered. It is worth noting that AM even-layer system does not satisfy the symmetry conditions required for the coexistence of spin splitting and spin degeneracy in AM odd-layer system mentioned above because the overall $\mathcal{M}_L$ symmetry for entire AM even-layer system is naturally satisfied if the adjacent interlayer $\mathcal{M}_L$ symmetry is satisfied. As shown in Figs. 5(e) and S9(b), the $Cr_2SSe$ quadlayer with AE-1 and AE-2 configurations exhibit overall $\mathcal{M}_L$ symmetry, and the completely compensated spin sublattices result in doubly spin degenerate electronic band (AFM-0). We note that in AM even-layer, regardless of whether adjacent interlayer $\mathcal{M}_L$ symmetry is present, the existence of overall $\mathcal{M}_L$ symmetry ensures complete spin degeneracy, which contrasts with the spin-layer coupling behavior in AM odd-layer that lack overall $\mathcal{M}_L$ symmetry. The external perpendicular gate field ($E_z$) further break the $\mathcal{M}_L$ symmetry for entire system, and the distinct spin splitting can be found in corresponding band structures (see Figs. 5(f) and S9(c)). Compared with conventional AFM multilayer, the spin/valley properties of AM multilayer with joint symmetry are highly regulated by magnetic configurations, stacking orders and external perpendicular gate field ($E_z$), and closely related to the spin-layer coupling, which is expected to be promising for spintronic device design.

## IV. CONCLUTION

In summary, we propose a design principle of spin-layer coupling in AM multilayer by first-principles calculations. From the perspective of spin group symmetry, the design principle of spin-layer coupling is demonstrated in Cr$_2$SX (X=O, S, Se) multilayer. The Cr$_2$SX monolayer possesses spin-valley locking composed of spin-polarized valleys, and the spin-valley splitting can be induced by breaking $\mathcal{M}_\Phi$ symmetry with uniaxial strain. The spin group symmetry in AM multilayer can be reduced by manipulation of magnetic orders, stacking configurations and perpendicular gate field, which is highly correlated with layer DOF to achieve nonrelativistic spin-layer coupling. Interestingly, the coexistence of spin splitting and spin degenerate is realized in Cr$_2$S$_2$ trilayer/pentalayer with type-1 configuration, which originate from the breaking of overall $\mathcal{M}_L$ symmetry, the presence of $\mathcal{M}_L$ symmetry in adjacent sublayer and $\mathcal{M}_z$ symmetry for entire system. Compared to AM odd-layer, the priority of overall $\mathcal{M}_L$ symmetry in AM even-layer is higher than that of interlayer $\mathcal{M}_L$ symmetry, which results in strong odd/even-layer dependence of spin-layer coupling. The perpendicular gate field ($E_z$) can directly couple the layer DOF to achieve the purpose of regulating the spin-layer coupling in AM multilayer. The symmetry analysis and spin-layer coupling principle can be readily extended to other members of AM, such as V$_2$Se$_2$O, V$_2$SSeO, V$_2$SeTeO and Cr$_2$O$_2$ monolayer and multilayer systems. Our work offers not only a series of 2D AM but also an efficient design principle to manipulate spin-layer coupling and spin/valley-related properties.

## ACKNOWLEDGMENTS

This work was supported by the Natural Science Foundation of Hebei Province (No. A2020202010).